\documentclass{emulateaph1}
\usepackage{graphicx}
\begin{document}
\bibliographystyle{apj}

\title{Diagnosing the Prominence-Cavity Connection}
\shorttitle{Cavity Emission Dynamics}

\author{Donald J. Schmit \altaffilmark{1,2}, Sarah Gibson \altaffilmark{1}}
\altaffiltext{1}{High Altitude Observatory, National Center for Atmospheric Research}
\altaffiltext{2}{Max Planck Institute for Solar System Research}
\shortauthors{Schmit and Gibson}

\begin{abstract}
Prominences and cavities are ubiquitously observed together, but the physical link between these disparate structures has not been established.
We address this issue by using dynamic emission in the extreme ultraviolet (EUV) to probe the connections of these structures.
The SDO/AIA observations show that the cavity exhibits excessive emission variability compared to the surrounding quiet-sun streamer, particularly in the 171\AA~bandpass.
We find that this dynamic emission takes the form of coherent loop-like brightening structures which emanate from the prominence into the central cavity.
The geometry of these structures, dubbed prominence horns, generally mimics the curvature of the cavity boundary.
We use a space-time statistical analysis of two cavities in multiple AIA bandpasses to constrain the energetic properties of 45 horns.
In general, we find there is a positive correlation between the light curves of the horns in the 171\AA~and 193\AA~bandpasses, however the 193\AA~emission is a factor of 5 weaker.
There is also a strong correlation between structural changes to the prominence as viewed in the He II 304\AA~bandpass and the enhanced 171\AA~emission.
In that bandpass, the prominence appears to extend several megameters along the 171\AA~horn where we observe co-spatial, co-temporal 304\AA~and 171\AA~emission dynamics.
We present these observations as evidence of the magnetic and energetic connection between the prominence and the cavity.
Further modeling work is necessary to explain the physical source and consequences of these events, particularly in the context of the traditional paradigm: the cavity is under dense because it supplies mass to the over dense prominence.
\end{abstract}
\setcounter{equation}{0}

\section{Introduction}
Prominences are regions of cool, condensed plasma suspended in the solar corona.
Our understanding of prominence physics has been divided into two distinct focuses: the topology of the magnetic support within the prominence and the mass source for the prominence itself.
Through the many iterations of experiments, there has been an overlooked constraint on the prominence system, coronal cavities.
Coronal cavities are the elliptical regions of hot (approximately 1-2 MK) plasma which surround the prominence.
Cavities have been observed to be density depleted as compared to the surrounding corona.
Rudimentary calculations, based on cavity volume and density, seem to indicate that the mass reduction in the cavity is not enough mass to populate the prominence \citep{saitotand_73}.
\\\indent
An important aspect of the cavity is the strongly contrasted boundary that is observed between the cavity and the surrounding streamer.
In the magnetically dominated corona, strong gradients in plasma properties (that would produce gradients in emission) must be related to the magnetic field and more specifically to distinct flux systems.
How is the cavity boundary related to the presence of stressed fields?
This question is central to our approach to studying the prominence-cavity system.\\\indent
Much research has been conducted attempting to diagnose the properties of the cavity.
\cite{gibson_10} presents a multiwavelength morphology study and contains the most comprehensive reference list.
Cavities were originally seen in white light scattering \citep{waldmeier_70,fuller_09} and these data measure unequivocally that cavities are a density-depleted feature.
Studies in X-ray and EUV observations \citep{serio_78} have found that cavities do emit at coronal temperatures.
Later studies in those spectral regimes uncovered that substructure was common in cavities \citep{hudson_99, reeves_12}, which raises the question: what causes the thermodynamics properties of the cavity to vary on smaller scales?\\\indent
In this article, we detail how dynamic EUV emission in the cavity can be used to extract structural and energetic information of the prominence-cavity superstructure. The layout of the article is as follows: Section 2 describes  generalized observations of dynamics in prominences and cavities, Section 3 characterizes the emission of prominence horns, and Section 4 discusses the conclusions we are able to draw from these observations as a whole.
\subsection{Prominence-Cavity Dynamics}
Prominences have long been observed to be dynamic phenomena \citep{rudnick34}.
The seminal observations are described in \citep{zirker}.
Doppler measurements are used to show that bulk motions in the prominence can be aligned but anti-parallel despite spatial thread separations of only few hundred kilometers (thread widths are on the order of the resolution element~$\approx100$ km).
This implies a highly coherent magnetic field but with distinctly non-coherent thermodynamic processes.
This study did not include any coronal datasets.
Higher resolution observations show that in addition to thread-motions, bubbles are also a dynamic prominence feature \citep{berger11}.
Bubbles are believed to be low density, magnetically-isolated structures which rise through prominence material buoyantly.
The prominence material passively shifts around the bubbles.
Both bubbles and thread motions operate on sub-megameter spatial scales, but despite these small scale dynamics, the large scale prominence structure remains intact.
Although, both of  these observations highlight the complex interaction of the magnetic field and plasma in the prominence, they both neglect to consider the larger role the cavity plays in the dynamics. \\\indent
The first observations of Doppler flows in cavities was reported in \cite{schmit_09}.
The coronal observations detailed in that article lacked the spatial and temporal resolution of the prominence observations.
However, it did find large-scale, long-duration flows to be present and common in cavities.
The source of these flows was not determined, but more importantly, the direct conclusion is that cavities which are observed to stably exist for weeks must also be regarded as a structure with internal dynamics, analogous to the prominence. \\\indent
To extend this work, we have observed cavities in the Solar Dynamics Observatory Atmospheric Imaging Assembly \citep[SDO/AIA]{lemen_12} dataset.
Although, the AIA dataset lacks spectral information to derive Doppler flows, it makes use of high resolution, high cadence bandpass imagers which span a broad range of temperature sensitivities.
With this dataset, we can observe the changes in emission which are indicative of changes in the thermodynamics properties of the prominence-cavity plasma.
Towards this aim, we are interested in one particularly relevant paradigm: is a mass exchange responsible for moving plasma from the under-dense cavity into the over-dense prominence?
There are various methods to describe this process but their commonality lies in thermal instability \citep{field_65, meerson_96}.
A coronal loop is perturbed such that a segment of the loop radiates more energy than the combination of heating and conduction input into the segment.
This causes a runway cooling effect in which the coronal plasma condenses as it cools to reach a new thermal equilibrium in a chromosphere-like condensation state.
Catastrophic cooling has been related to two different processes: shearing field lines \citep{choe_92} and chromospheric heating \citep{karpen_06}.\\\indent
This article focuses on diagnosing the dynamics of the prominence-cavity.
We establish with this article that there is a magnetic and energetic connection between the prominence and the cavity.
A subsequent article by these authors will compare the observed dynamics to the hydrodynamic models of catastrophic cooling.
This followup article will address the physical source and consequence of that energetic connection.
\section{EUV Dynamics in the Prominence Cavity}
We use the SDO/AIA dataset to observe dynamics in the prominence-cavity system in the EUV.
SDO/AIA images the Sun in eight EUV and UV bandpasses which are centered on strong emission lines.
The strength of these emission lines vary with temperature so the AIA datasets present us with a qualitative temperature diagnostic tool.
Of the 8 bandpasses on AIA, four of the bandpasses are useful for prominence-cavity diagnostics: 304\AA, 171\AA, 193\AA, and 211\AA.\\\indent
The 304\AA~bandpass is centered on the singly-ionized Helium Ly-$\alpha$ transition, although it also includes a Si XI line which is generally two orders of magnitudes dimmer.
He II 304\AA~emission is believed to primarily occur between $0.5-1.0\times10^5$K although extended optically thin emission throughout the transition region is also possible.
The 171\AA~bandpass is centered over a blend of Fe IX and Fe X emission which is strongest at a temperature of 1 MK.
These are the strongest lines by an order of magnitude but there is a weak transition region component between 2.5-5$\times10^5$ K formed by O V-IV.
The 193\AA~bandpass centers on an Fe XII transition which is characteristic of 1.5 MK plasma.
The  211\AA~bandpass is centered on a Fe XIV transition at 211.3\AA.
In ionization equilibrium, Fe XIV reaches a population maximum at 2 MK.
In terms of spectral blends, the 211\AA~bandpass is fairly pure, but its diagnostic value is damped by its signal strength.
\\\indent
SDO/AIA allows us to image the solar plasmas with the temperature range of 0.05-2 MK.
In practice, the observations provide a convolution of many different temperature structures which complicates our interpretation.
Moreover, the data contains instrumental effects and observational uncertainties, which we must derive and understand to accurately state our measurements.\\\indent
Throughout this section, we will be discussing two particular prominence-cavity datasets.
The first is a 30-hour dataset collected between 28-Feb-2011 09:00UT and 1-Mar-2011 15:00UT, hereby known as D1.
The cavity of interest is off the northwest limb at a latitude of 60$^\circ$.
Based on the duration of visibility, the longitudinal length is estimated at 52$^\circ$ or 320 Mm.
The second is a 36-hour dataset collected between 10-Aug-2011 00:00UT and 11-Aug-2011 12:00UT, hereby known as D2.
This cavity is off the southeast limb at a latitude of -51$^\circ$.
The longitudinal length is estimated at 38$^\circ$ or 340 Mm.
Each dataset is composed of full resolution data at 6 minute cadence.
Each image in the dataset is the result of median filtering 5 sequential images to increase signal-to-noise ratio.\\\indent
The first steps to understanding the role of dynamics in cavities is to understand statistically how the cavity differs from the surrounding streamer.
\subsection{Characterization of Prominence-Cavity Structure in EUV}
\begin{figure*}
\center
\includegraphics[width=0.8\textwidth]{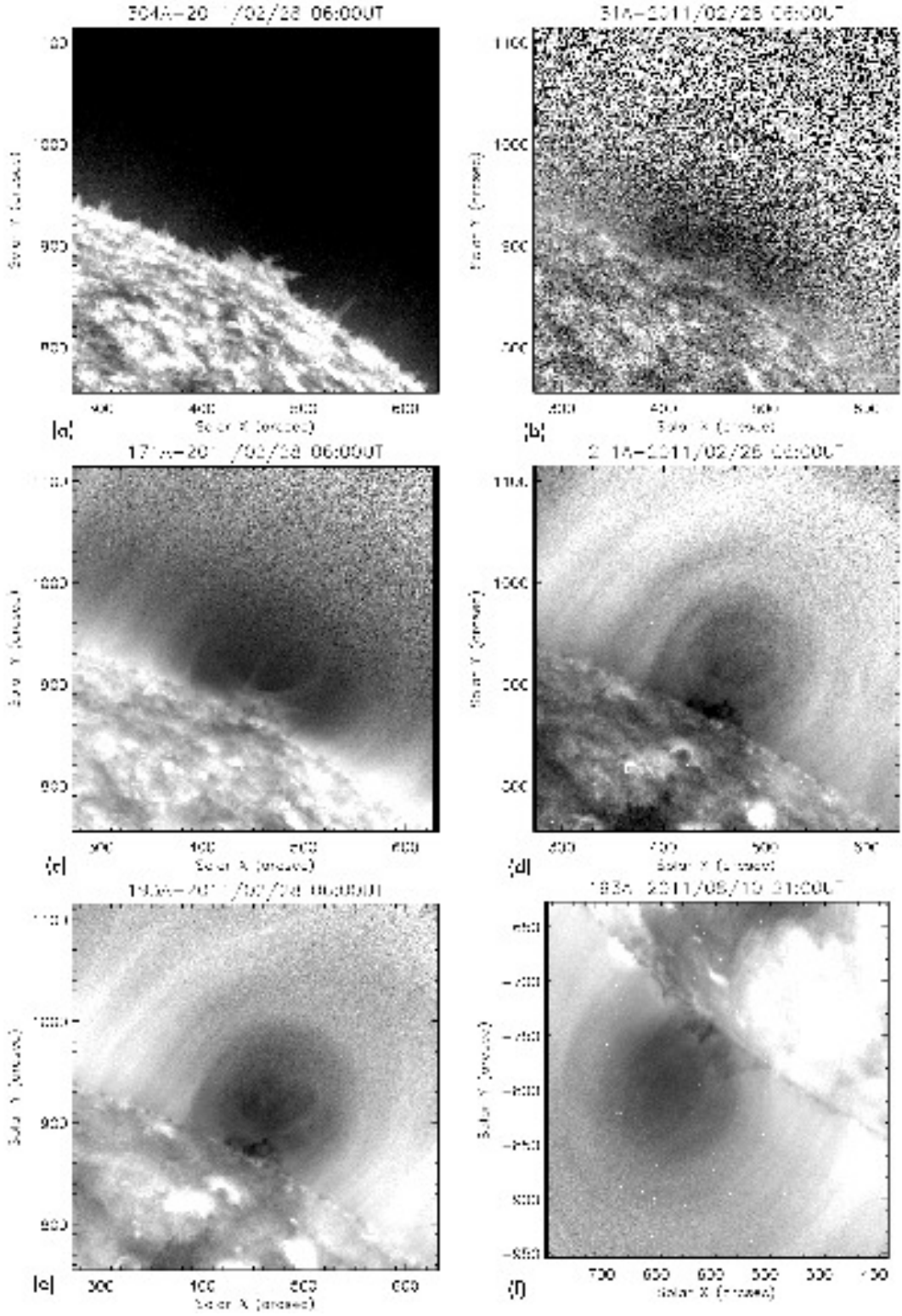}
\caption{Maps in several SDO/AIA bandpasses of cavity D1 (a-e) and D2 (f). He II 304\AA~(a). Fe VIII/Fe XXIII 131\AA~(b), Fe IX/Fe X 171\AA~(c), Fe XIV 211\AA~(d), Fe XII 193\AA~of D1 (e) and D2 (f). Data have radially vignetted to promote structural contrast.\label{fig:mult}}
\end{figure*}

Figure~\ref{fig:mult} displays the D1 cavity in five AIA bandpasses.
The 304\AA~bandpass displays the cool, dense prominence.
This particular prominence extends 18 Mm above the limb and has an apparent width of 30 Mm.
There is no clear separation between the limb and lower boundary of the prominence.
\\\indent
Figure 1b shows the prominence in 131\AA.
The 131\AA~bandpass data theoretically contains emission from a FeVIII line (blended with the primary Fe XXIII line).
The 131\AA~bandpass simply does not have enough signal to aid in this analysis.\\\indent
The signal in the 171\AA~bandpass is much stronger than 131\AA.
The prominence appears in emission.
This emission is peculiar but ubiquitous in prominences.
The peak ionization temperature for Fe IX is 8 times higher than that of He II, but there is apparent cospatial 304\AA~and 171\AA~emission in the prominence.
The explanation for this phenomenon is the ``prominence-corona transition region'' (PCTR) \citep{orrall_76}.
The most basic interpretation is that along lines of sight that intersect the prominence, there is emission from coronal, transition region, and condensation plasma.
The 171\AA~emission we see in the prominence is being emitted by the transition region between condensed plasma and the corona.
\\\indent
The 193\AA~bandpass shows the cavity very clearly as an emission-depletion feature.
The coolest mass of the prominence can be seen in absorption, which is formed by bound-free continuum absorption mostly due to H I.
For reference, a 193\AA~image of D2 is also presented in Figure~\ref{fig:mult}f.
\\\indent
The cavities in D1 and D2 differ primarily in two ways.
First, D2 is centered around a more angled neutral line.
It appears to move north to south as the cavity rotates onto the solar disk.
This has the affect of diffusing the contrast between the cavity interior and the streamer.
Second, the streamer surrounding the cavity in D2 is brighter in the coronal bandpasses than the streamer in D1.
This can be seen more clearly on the solar disk, where the filament channel is significantly more contrasted than in D2.
\subsection{Temporal Statistics on Emission Variability}
\begin{figure*}
 \center
\includegraphics[width=0.8\textwidth]{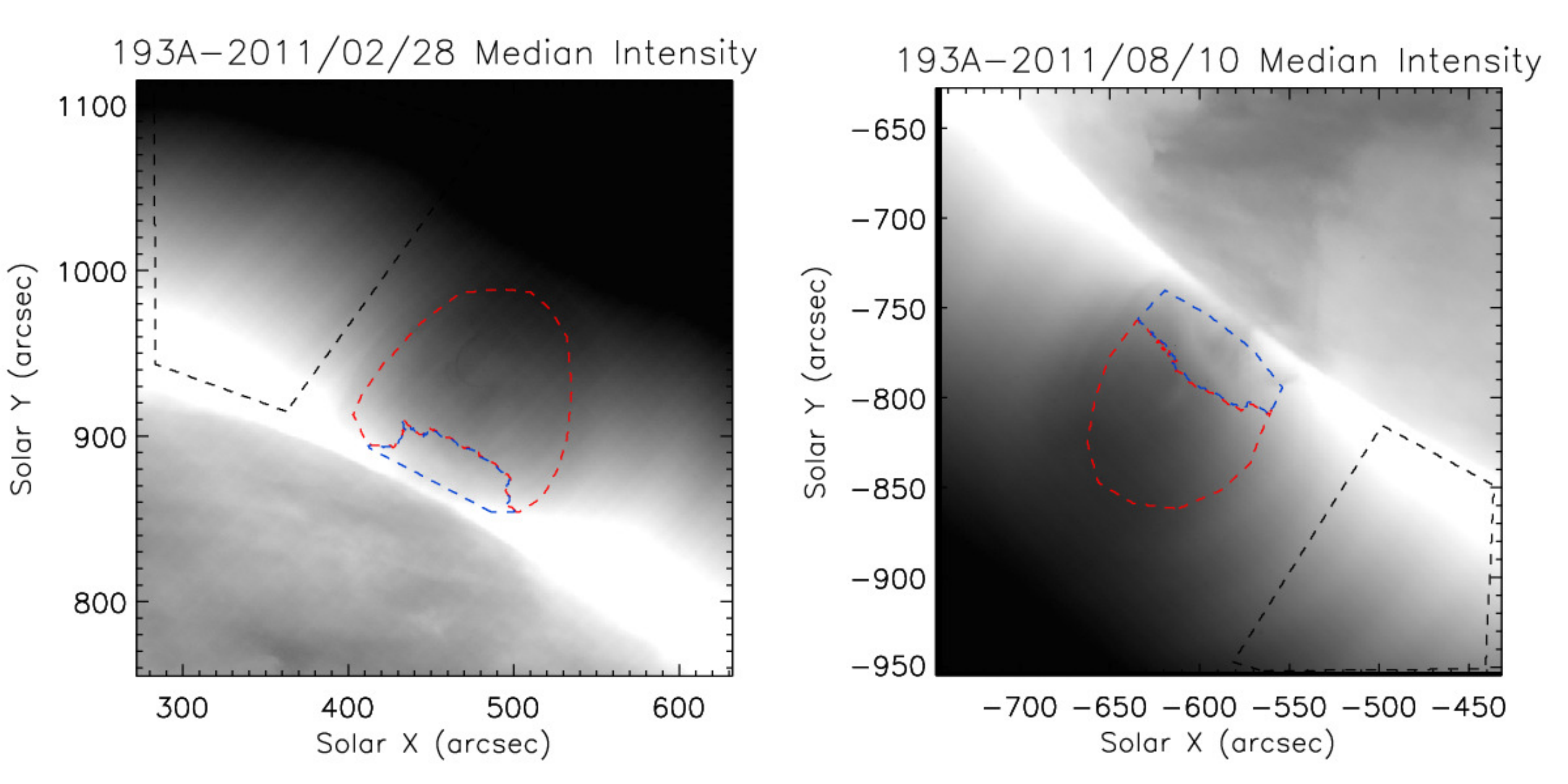}
\caption[Spatial regions of streamer, cavity, and prominence]{Time averaged maps in 193\AA~ for D1 (left) and D2 (right). The red dashed contour denotes the regions selected as cavity. Blue region is prominence. Black region is streamer. These color codes are used in Figures~\ref{fig:intvr} and~\ref{fig:vari} as well.\label{fig:reg}}
\end{figure*}

\begin{figure}
\center
\includegraphics[width=0.5\textwidth]{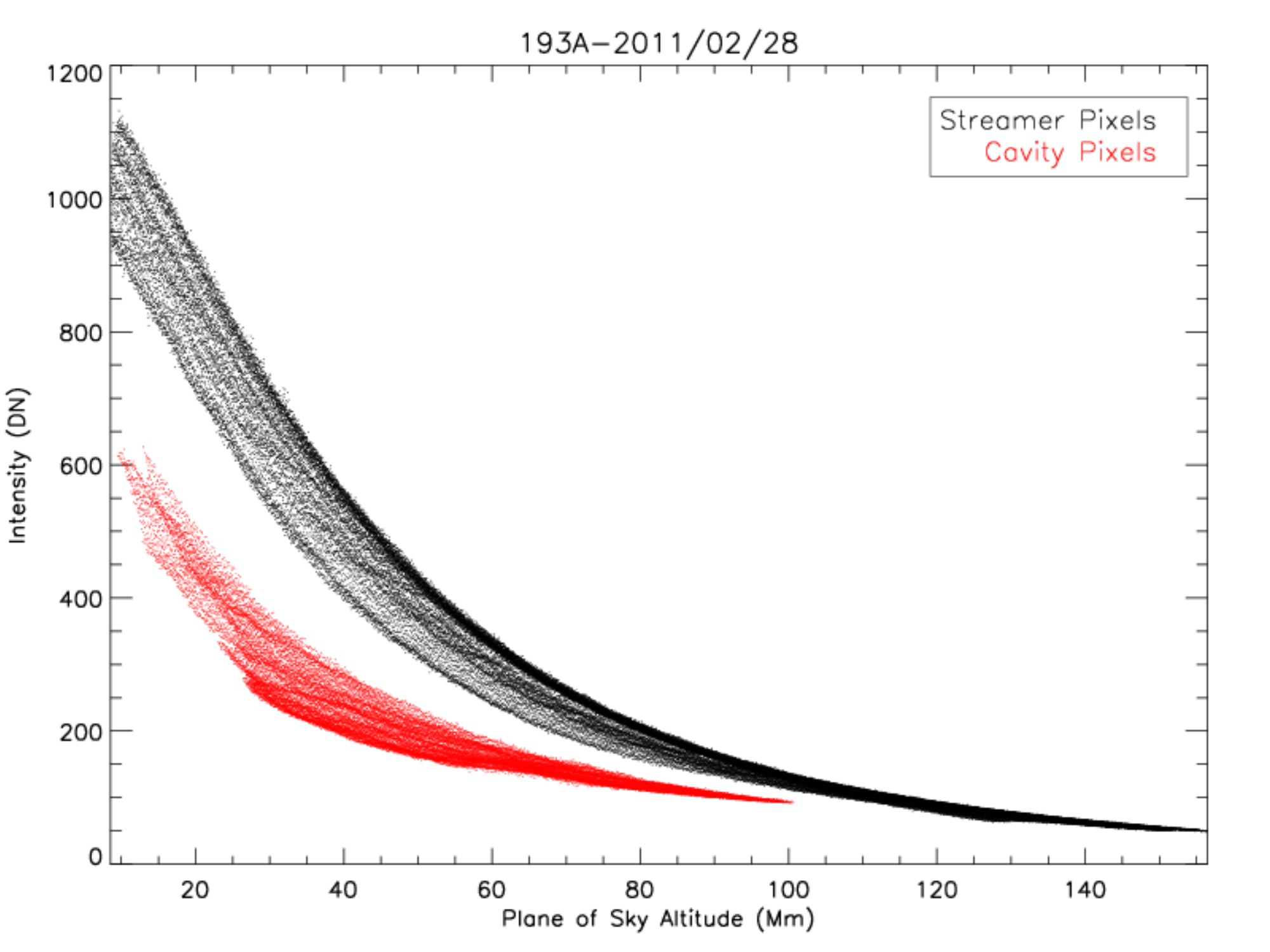}
\caption[193\AA~intensity depletion in the D1 cavity]{Radial profiles of the median intensity, $\mathcal{I}^{ij}(p=0.5)$, for D1 in 193\AA. As in Figure 2, red points are extracted from the cavity region, black points are extracted from the streamer region.\label{fig:intvr}}
\end{figure}

\begin{figure*}
\includegraphics[width=\textwidth]{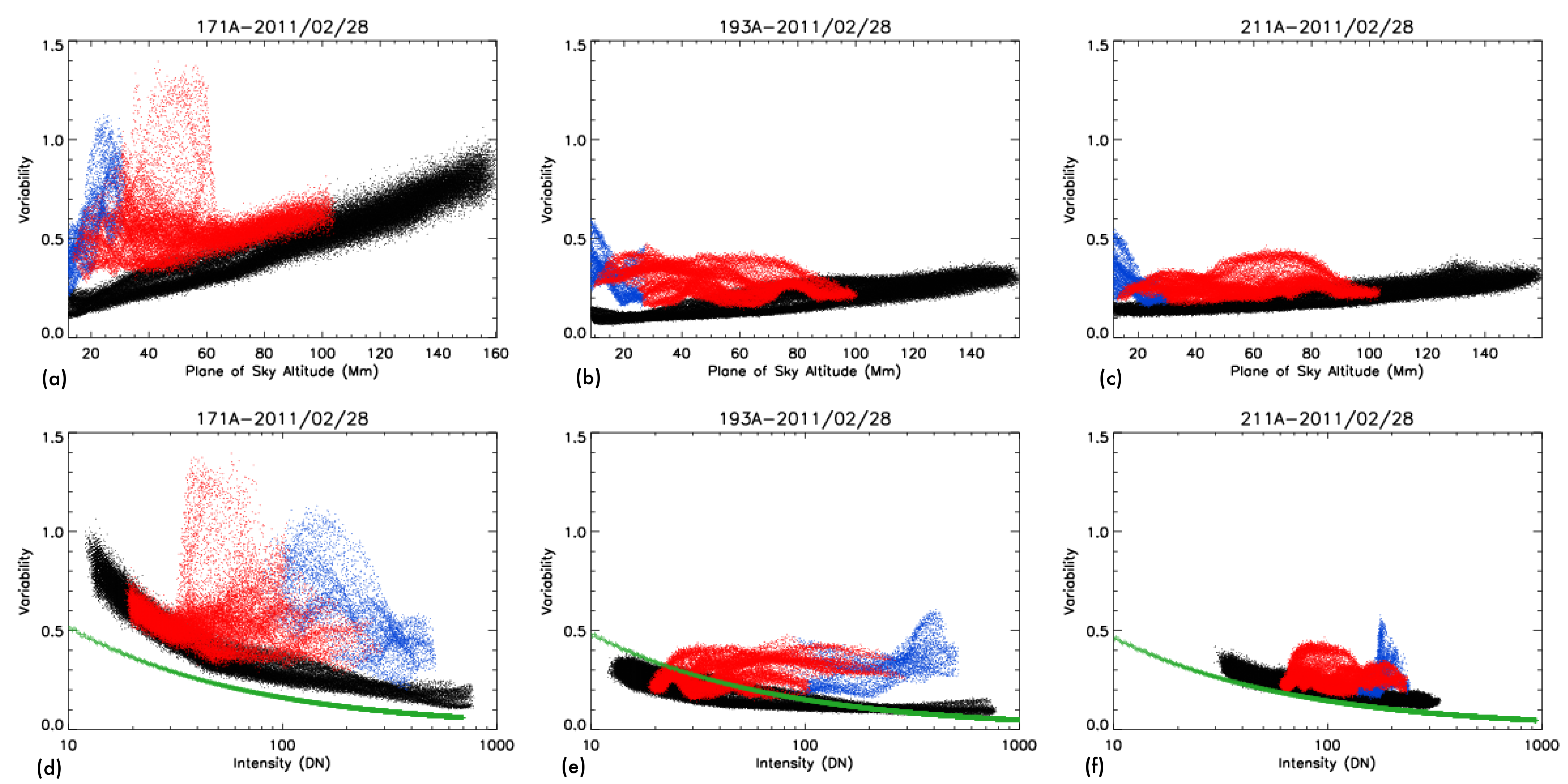}
\caption[Structural dependence of the variability statistic]{Variability statistic plotted against radial height (a-c) and against intensity (d-f) for D1 in 3 different bandpasses: 171\AA~(a and d), 193\AA~(b and e), 211\AA~(c and f). The green profiles are the theoretical Poisson-noise variability. Otherwise, the color code is based on the regions of Figure 2. \label{fig:vari}}
\end{figure*}

Figure~\ref{fig:mult} served to qualitatively present the morphological differences of the prominence cavity in the various AIA bandpasses.
We will now develop a set of statistics to quantitatively describe how the time-dependent cavity differs from the streamer.
The AIA observations can be described in the following notation:
\begin{displaymath}
^\lambda{\bf I}_t^{ij}
\end{displaymath} 
where {\bf I} is the observed intensity (in data number) in bandpass $\lambda$ at time $t$ for CCD pixel $ij$.
We are attempting to understand the time dependent plasma processes ongoing in the cavity and how those properties differ from the streamer, thus we will largely be ignoring the spatial coordinates $ij$.
Instead we will subdivide the dataset into regions: streamer, cavity, and prominence.
The subdivided regions are shown in Figure~\ref{fig:reg}.
Figure~\ref{fig:intvr} illustrates how the median intensity (in time at each individual pixel) in 193\AA~varies between the streamer and the cavity as a function of radial height in the plane of the sky.
The cavity data is presented as red points, while the streamer data is presented as black points.
The cavity profile is everywhere reduced compared to the streamer at the same height.
This emission depletion varies between 20-50\%.
Assuming an isothermal corona and ignoring projection effects, the $\int n^2 d\ell$ emission relationship would imply a 29\% density depletion.
This is typical of cavities as reported by \cite{fuller_09}.
To analyze the time-dependent variability in the AIA data, we will use statistics based on the time series of intensity at every pixel within the dataset.
\\\indent We have created a statistic, $V^{ij}$ referred to as variability, designed to quantify the degree to which the intensity of a given pixel varies as a function of time.
Variability is given by
\begin{equation}
V^{ij}=\frac{I^{ij}_{MAX}-I^{ij}_{MIN}}{I^{ij}_{MED}}
\end{equation}
where $I^{ij}_{MIN}$, $I^{ij}_{MAX}$, and $I^{ij}_{MED}$  represent the intensity value that pixel $ij$ is brighter than through 95\%, 5\%, and 50\% of the time series, respectively ($I^{ij}_{MED}$ will be referred to as median intensity).
This statistic is akin to the variance, however the nature of the intensity distributions (i.e. asymmetric) does not make variance a particularly illustrative statistic.
Whereas variance weights the entirety of a distribution, our variability statistic cares specifically about the ``wings'' of the intensity histograms.
It measures how the intensity span of the wings compares with median intensity.\\\indent
Figure~\ref{fig:vari} plots variability in the three structural regions for 171\AA, 193\AA, and 211\AA.
The upper row of plots presents variability versus radial height.
Examining the streamer distributions for the three bandpasses, we find there is a linear relationship between variability and altitude for all bandpasses.
Moreover, the distributions are fairly tightly clustered.
The cavity and prominence distributions are not so simple.
The 193\AA~and 211\AA~data both show variabilities which are stronger in the cavity than the streamer.
The prominence shows higher variability than the cavity at low heights.
In the 171\AA~distributions, there is a striking difference between the cavity region and the streamer region.
The cavity has a variability that is a factor of 4 higher than the variability of the streamer.
The differences between the variability distributions illustrate that the cavity is much more variable than the streamer, but specifically in the cooler 171\AA~bandpass.
\\\indent The lower row of Figure~\ref{fig:vari} presents the variability data as a function of median intensity.
While the upper row illustrated that the cavity is more variable than the streamer, it does not describe the significance of that variability.
To define the significance of the statistic, we must define a baseline for what the expected observational error should be.
With the AIA data, this is challenging for two reasons.
First, due to the volume of data, the AIA datasets are only retrievable as ``Level 1'' data so the engineering level data (flat field and dark current) are not available, although these systematics have been accounted for at Level 0 to Level 1 conversion.
Additionally, bandpass imagers cannot photon count because there is an indefinite range of photon energies impinging on the detector.
Due to the non-uniform energy response of the detector, this does not allow a singular photon to electron to data number gain to be assigned.
In the lower plots, the expected Poisson noise variability based on the preflight detector calibration is plotted as a function of intensity as green points.
In general, the Poisson variability is a close match to the streamer distributions.
The 211\AA~data is perhaps the clearest example of this.
The offset between the streamer distribution and Poisson noise can be attributed to a small variability in emission.
\\\indent
The most important feature to take away from Figure~\ref{fig:vari} is that the variability of the cavity in 171\AA~is excessively strong compared to both the streamer and the baseline Poisson noise.
There is also excessive variability in the cavity compared to the streamer in 193\AA~and 211\AA, but the 171\AA~data is a factor of 3 more variable.
Through this statistical analysis, we have identified a unique characteristic of the EUV emission in the cavity through the basis of spatially-independent temporal statistics.
Now, we will take the next step by honing in on the strongly variable regions of 171\AA~cavity.
What is it about the intensity distributions in the cavity region that increases this statistic?
\begin{figure*}
\includegraphics[width=\textwidth]{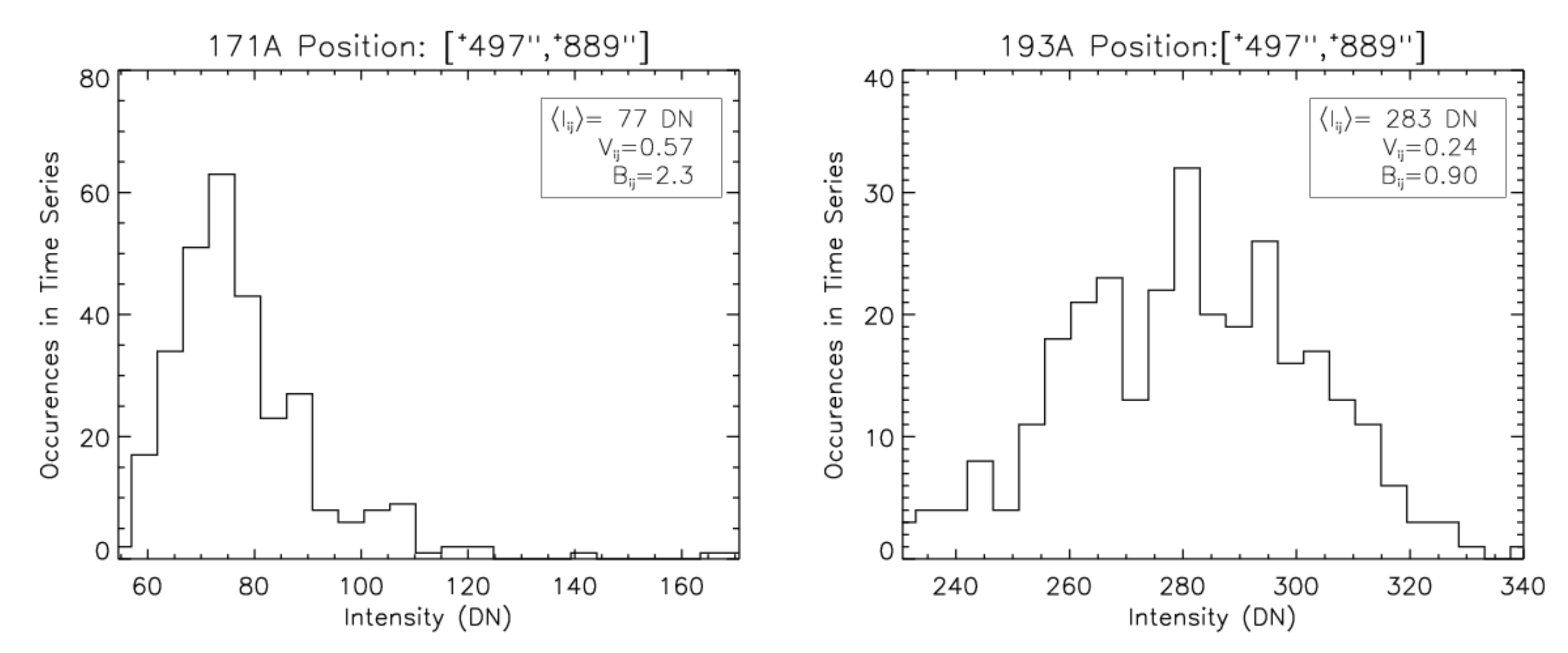}
\caption[Typical intensity histograms in the cavity]{Intensity histograms for the intensity of a cavity region pixel of D1 in the 171\AA~bandpass (left) and the 193\AA~bandpass (right).\label{fig:hist}}
\end{figure*}
\\\indent Figure~\ref{fig:hist} shows the histogram for a strongly variable cavity point in the 171\AA~and 193\AA~bandpasses.
The variability in the 171\AA~data is over twice that in 193\AA.
In terms of the distributions, the most notable feature is the strong asymmetry in the 171\AA~distribution, which has a high intensity tail.
The 193\AA~histogram, in contrast has a symmetric normal distribution.
We have chosen statistics to highlight the nature of these distributions, as the intensity histograms plotted in Figure~\ref{fig:hist} are typical of the cavity.
The high intensity wing in the 171\AA~histogram make both mean and variance (statistics of specific relevance in a normal distribution) of little illustrative value.
Printed along with the variability statistic in Figure~\ref{fig:hist} is the skew statistic:
\begin{equation}
 B^{ij}=\frac{I^{ij}_{MAX}-I^{ij}_{MED}}{I^{ij}_{MED}-I^{ij}_{MIN}}.
\end{equation}
Whereas the variability looks at the spread between the high-intensity and low-intensity wings of the intensity histograms, the skew statistic looks at the relative intensity-difference between congruent positions in the histogram, specifically the brightest 5\% of intensities and the dimmest 5\% of intensities.
The skew is a positive number, and for the AIA datasets it varies between 0.1 and 10.
A skew of 1 would imply a perfectly symmetric distribution.
We find the lower cavity is characterized by high skew distributions ($B_{ij}>2$) in the 171\AA~bandpass but not 193\AA~bandpass.
This signifies that the variability in the 171\AA~cavity is characterized by short duration, strong brightening events.
\\\indent
Based on the skew and variability statistics, we have found that the cavity exhibits very different temporal variations in intensity than the streamer.
Based on these statistics, we now inspect movies of the cavity to determine the spatial and temporal coherency of the 171\AA~brightening events.
\section{Observations of Prominence Horns}
\begin{figure*}
\includegraphics[width=\textwidth]{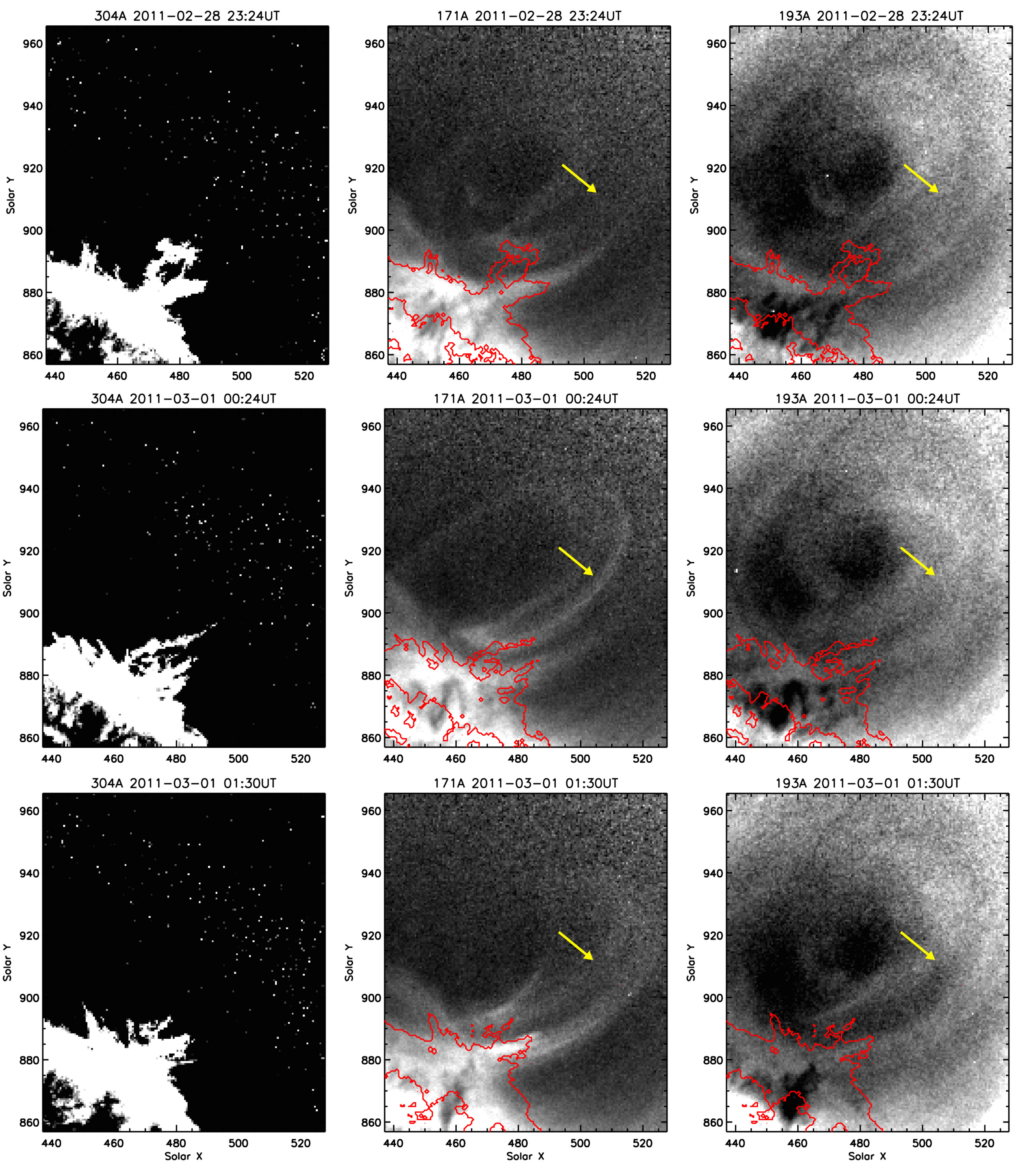}
\caption[Prominence horn in the cavity as seen in AIA]{Snapshots at three times in three bandpasses of a prominence horn, zoomed into D1. Prominence material, He II 304\AA~emission (left). Transition region Fe IX-X 171\AA~(middle). Coronal Fe XII 193\AA~(right). The horn is the 171\AA~brightening which the yellow arrow points to. It remains nominally brighter than the background for 1.5 hours.  The red contours show the perimeter of strong 304\AA~emission. The horn in co-spatial with an extension of the prominence. There is weak, correlated signal in 193\AA~radially above the yellow arrow, but overall the structure in 193\AA~is distinct from that of 171\AA. A movie of this figure is presented in the electronic version.\label{fig:close} }
\end{figure*}

By examining movies of the 171\AA~bandpass data, the nature of these variability features become more apparent.
Figure~\ref{fig:close} illustrates a subregion of the cavity going through a transient 171\AA~brightening, and an animated version of this figure is presented in the electronic edition of this article.
The pictured subregion includes the prominence and the western half of the cavity.
Over the course of an hour, we observe the formation of a collimated ``horn'' in the 171\AA~data which extends nonradially from the 304\AA~prominence into the lower cavity.
The apparent width of the feature is 4.2 Mm and the apparent length (in the plane of the sky) is 63 Mm.
The maximum brightening (contrast relative to background) is 16\%.
The spatial geometry of the feature extends from the prominence along a concave up loop-segment into the cavity.\\\indent
The correlation between the 171\AA~feature and the structure of the other AIA bandpasses should also be considered.
In between 2011-02-28 23:24UT and 2011-03-01 00:24UT, we find that the prominence emission, as seen in the 304\AA~bandpass, has also changed.
A co-spatial, co-temporal extension of the prominence sits at the base of the 171\AA~horn.
The prominence gradually extends a maximum of 13 Mm along the 171\AA~feature before gradually contracting.\\\indent
This observation demonstrates that there is a coronal component associated with prominence dynamics, and moreover the dynamic coronal feature projects into the cavity interior.
These correlated features imply there is a magnetic and energetic link between the corona and prominence, which leads us to the question: what is the physical source for the prominence-corona dynamics?
To address this question, we will look towards statistical analysis of the many prominence horns to identify physical characteristics which will aid us in determining the cause and consequence of these features.
\subsection{Method of Analyzing Prominence Horns}
The first step in this analysis is to build a database of events.
As we previously established, the most dominant forms of emission variability in the cavity are collimated regions of strong brightening in 171\AA~emission.
We use this trait as the defining characteristic of feature identification, and we use visual inspection to identify potential features in both D1 and D2.
We originally attempted to develop an automated method to extracting features through a wavelet approach similar to the OCCULT algorithm developed by \cite{aschwanden_10b}.
These techniques are not well suited to off-limb corona, where three effects hamper them: the strong radial intensity gradient, the relatively diffuse structure, and the lower signal-to-noise ratio.
Instead of investing time into automating the initial processing, we have opted to manually identify potential features, which are later verified through tests described later in this section.
We identify a total of 48 potential features (in 66 total observing hours) between our two datasets: 27 features in D2 and 21 features in D1. \\\indent
To quantify the changes in intensity as a function of time along each feature, we synthesize a curved slit which is selected to overlay the potential feature.
The data from the 193\AA, 171\AA, and 304\AA~datasets are extracted in the vicinity of the slit and are transformed into a curvilinear coordinate system, the axes being distance along the slit, $y'$, and distance from slit-center, $x'$.
The transformation is done using a triangulation, nearest-neighbor interpolation scheme.
An example of a transformed curvilinear dataset is shown in Figure~\ref{fig:curv}.
In the temporal domain, we extract data from 2 hours prior to the potential feature and 3 hours after.
\begin{figure}
\includegraphics[width=0.5\textwidth]{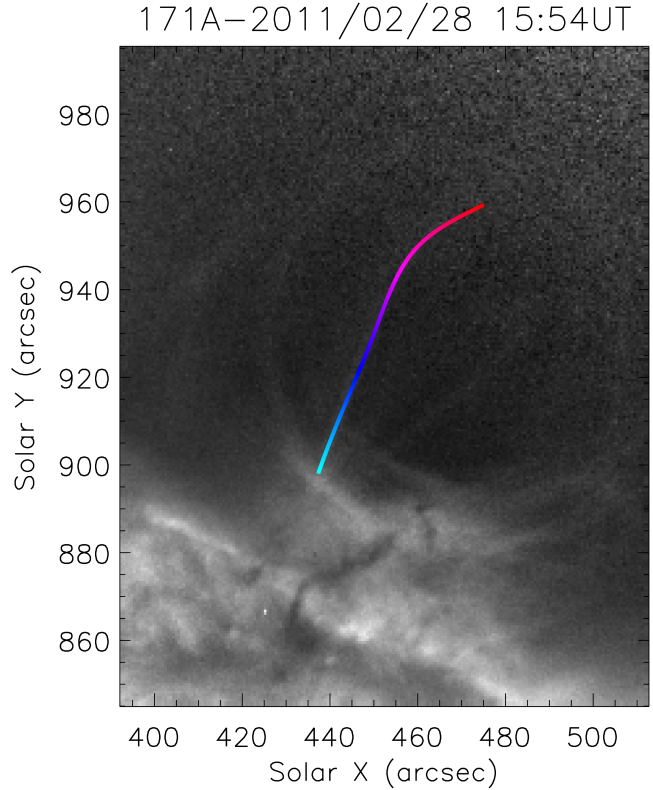}
\caption[Curvilinear horn datasets]{ A 171\AA~image overlaid with a curved slit. The color of the slit represents the curvilinear $y$-coordinate as shown in Figure 8.\label{fig:slit}}
\end{figure}
\begin{figure*}
\includegraphics[width=\textwidth]{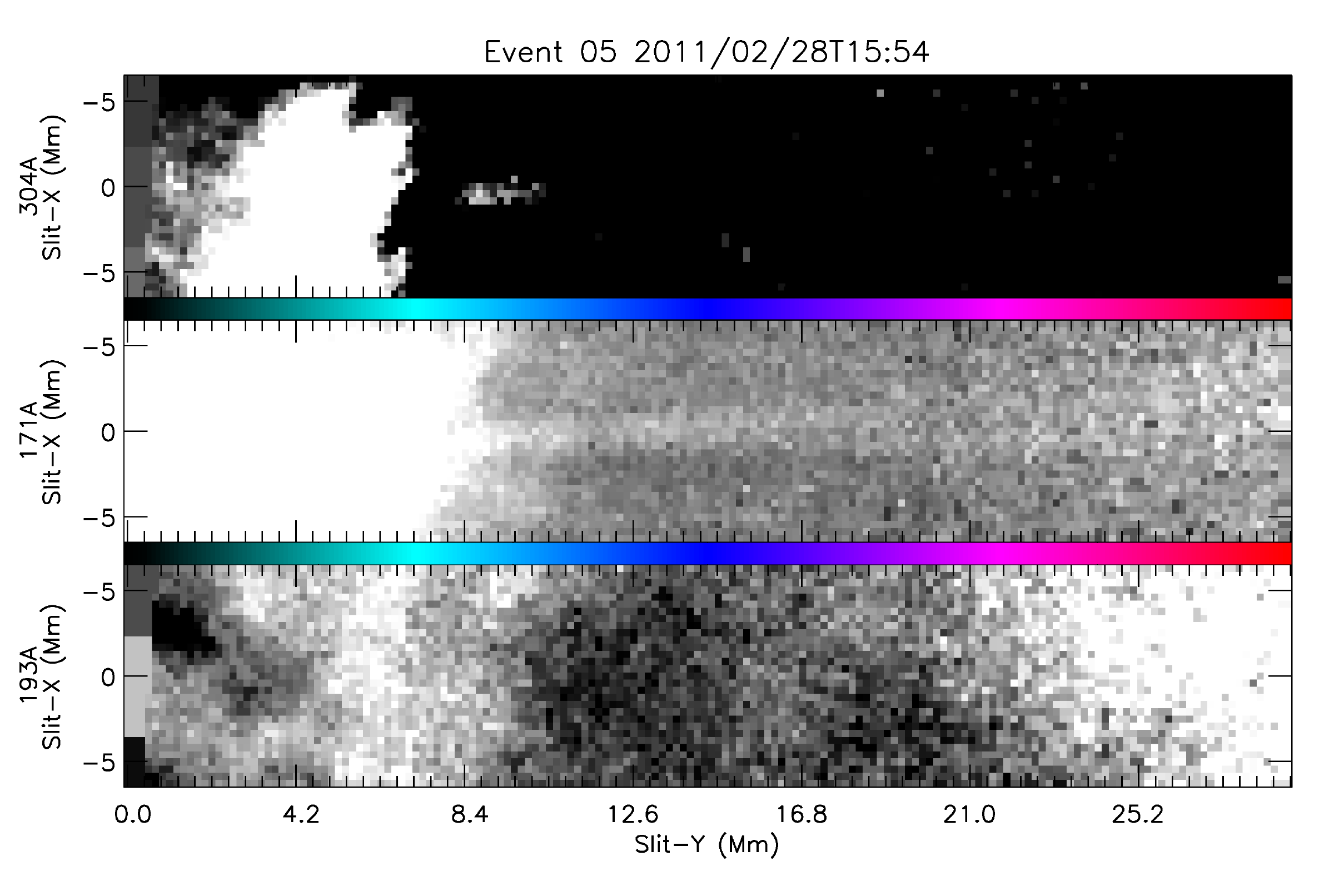}
\caption[Curvilinear horn datasets]{A single horn image, transformed into curvilinear coordinates using a curved slit.  In this event, a blob is ejected from the prominence to a slit-height of 9 Mm. This ejecta coincides with a coherent 171\AA~brightening which extends from 5 Mm to 23 Mm. There is also a correlated 193\AA~brightening which is not visible in this map because it has not been background-divided. The color table plotted horizontally between maps indicate the slit-height of the light curves which are plotted in Figure 9.\label{fig:curv}}
\end{figure*}
\\\indent In order to quantify the time-dependent intensity variations in these datasets, we produce light curves as a function of position along the slit.
The curvilinear dataset is divided into background regions (5 pixels$<|x'|\le 15$ pixels) and feature regions ($|x'|\le 5$ pixels).
These regions are then coadded at 5 pixel intervals in the $y'$-direction (along the slit).
A gaussian temporal smoothing is applied to the light curves, with a width of 9 minutes.
We present light curves which display the ratio of feature intensity to the background intensity at each segment along the slit as a function of time.
Example plots of these light curves are shown in Figure~\ref{fig:prof}.
The color of each curve denotes in position along the slit.
A colorbar matching the $y'$ coordinate is plotted in between maps in Figure~\ref{fig:curv}.
A vertical offset has been applied to the light curves shown in Figure~\ref{fig:prof} so that they do not lie on top of each other.
\\\indent
The final step in the processing of our datasets is finding and fitting the locations of minima and maxima of each individual light curve.
This process serves two purposes: it allows us to automatedly check to make sure that the potential features we identified by inspection are actually coherent structures, and it allows us to develop statistics on the onset, duration, and time-lag between different bandpasses of each event.
For each light curve, we use an algorithm to determine the location of local maxima between $0<t<40$ (where $t=20$ is the time of maximum feature definition).
For a potential feature to be labeled as an event, we require that at least 8 sequential heights along the slit ($\approx$17 Mm) have local temporal maxima within 3 timesteps of both the neighboring heights.
This basic test is a simple method which checks the coherency of a potential feature.
Applying this filter to our 48 potential features, we return 45 confirmed prominence horns. \\\indent
We extend this technique to measure the start and end times of each brightening.
These time intervals are assessed by extracting the nearest two local minima on either side of the prominence-horn maximum.
We apply a linear fit to these data (time of event vs. position on the slit) where the weight of each minima is inversely proportional to its depth relative to the maxima:
\begin{displaymath}
\sigma=(I(t=t_{MAX})-I(t=t_{MIN}))^{-1}
\end{displaymath}
where the linear fit takes the form
\begin{equation}
t'=m*y'+t_0.
\end{equation}
The quantity $t'$ is the best-fit time of event (event in this context being the initial brightening, peak brightness, end of subsequent dimming).
The derived quantity $m$ has units of km$^{-1}$ s.
Its reciprocal is therefore a velocity which gives us an estimate of how the observed disturbance propagates along the slit.
The fitting routine minimizes the quanity,
\begin{equation}
\chi^2=\sum_{y'} \frac{(t_{obs}(y')-t'(y'))^2}{\sigma^2}.
\end{equation}
where $t_{obs}$ is the observed timestep of the event.
The weighting function values deeper minima more than shallow minima, and strictly-based on number of points, minima which are well correlated between several heights result in lower $\chi^2$ values.
Figure~\ref{fig:prof} has black lines overplotting the derived linear fits for the feature initiation, peak, and dimming.
\begin{figure}
\includegraphics[width=0.5\textwidth]{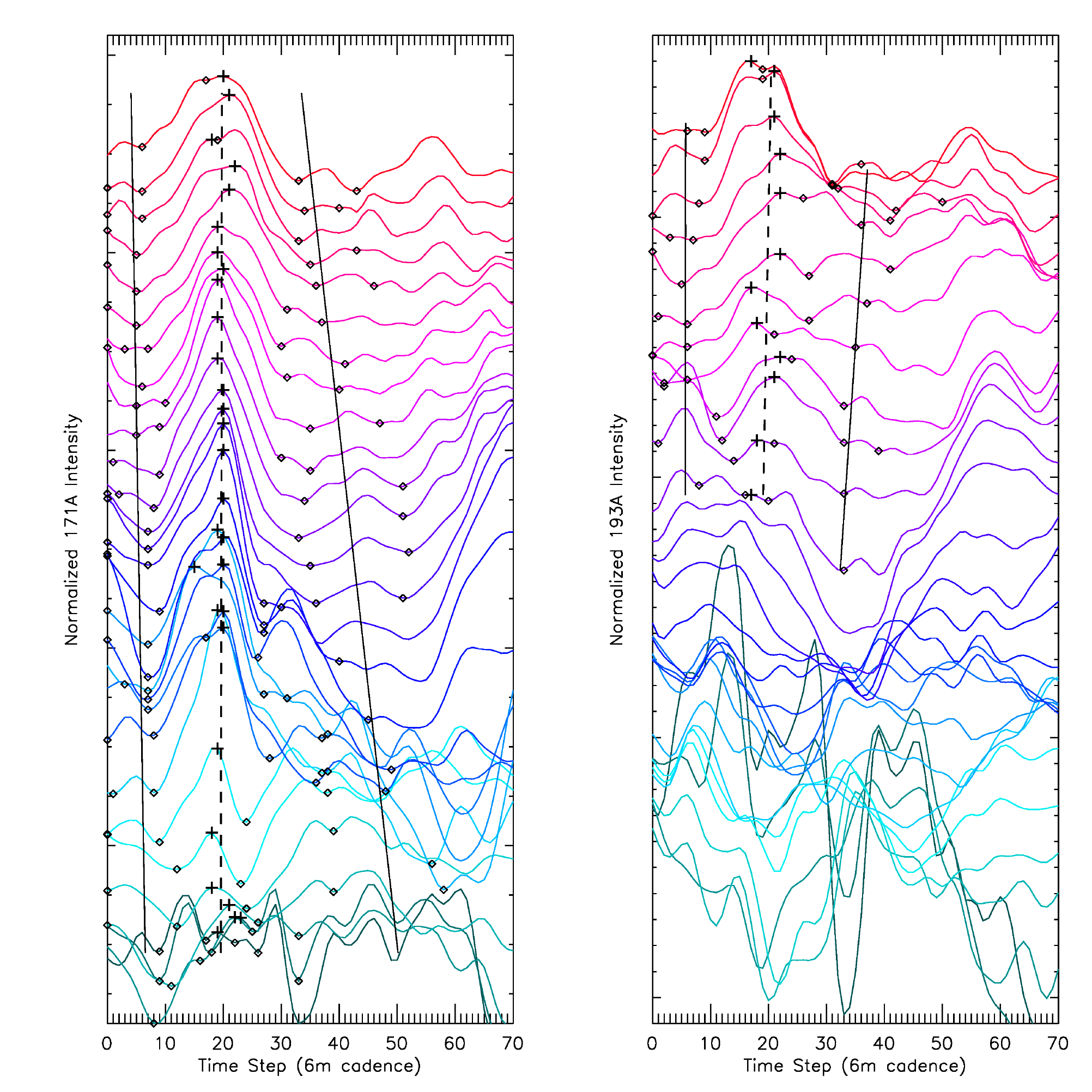}
\caption[Light curves from a prominence horn]{Light curves of Event 5 (shown in Figure 8) in the 171\AA~bandpass (left) and the 193\AA~bandpass (right). The color of the light curve denotes its height along the slit (see colorbar, Figure 8). The teal curve is extracted from a height of 4 Mm while red curve is extracted at a height of 27 Mm.  The light curves are offset in the vertical direction for clarity. The time of maximum brightness contrast of the feature is marked with black crosses, while the surrounding minima are marked in black diamonds. The linear fit (Equations 3) is overplotted with black lines to show the temporal progression of brightening event as a function of slit-height.  \label{fig:prof}}
\end{figure}
\subsection{Results}
Using the above described methodology, we will now go into characterization of prominence horns.
The basic configuration for 171\AA~events is as follows:
a rapid, large-scale brightening begins with its base near the prominence.
From onset time to peak brightness takes 85 minutes with a standard deviation of 19 minutes.
The brightness enhancement is not constant with distance along the feature, and
the brightest point along the feature tends to occur within the first 15 Mm.
The peak brightness contrast generally decreases as a function of distance along the horn, such that the horn fades into the background at higher heights.
We observe that there is a weak correlation (Pearson coefficient: 0.26) between the peak brightness of a horn and its length.
For the 45 measured horns, the median peak brightness contrast is 34\%, and 9 features have a peak brightness over 50\%.
\begin{figure*}
\includegraphics[width=\textwidth]{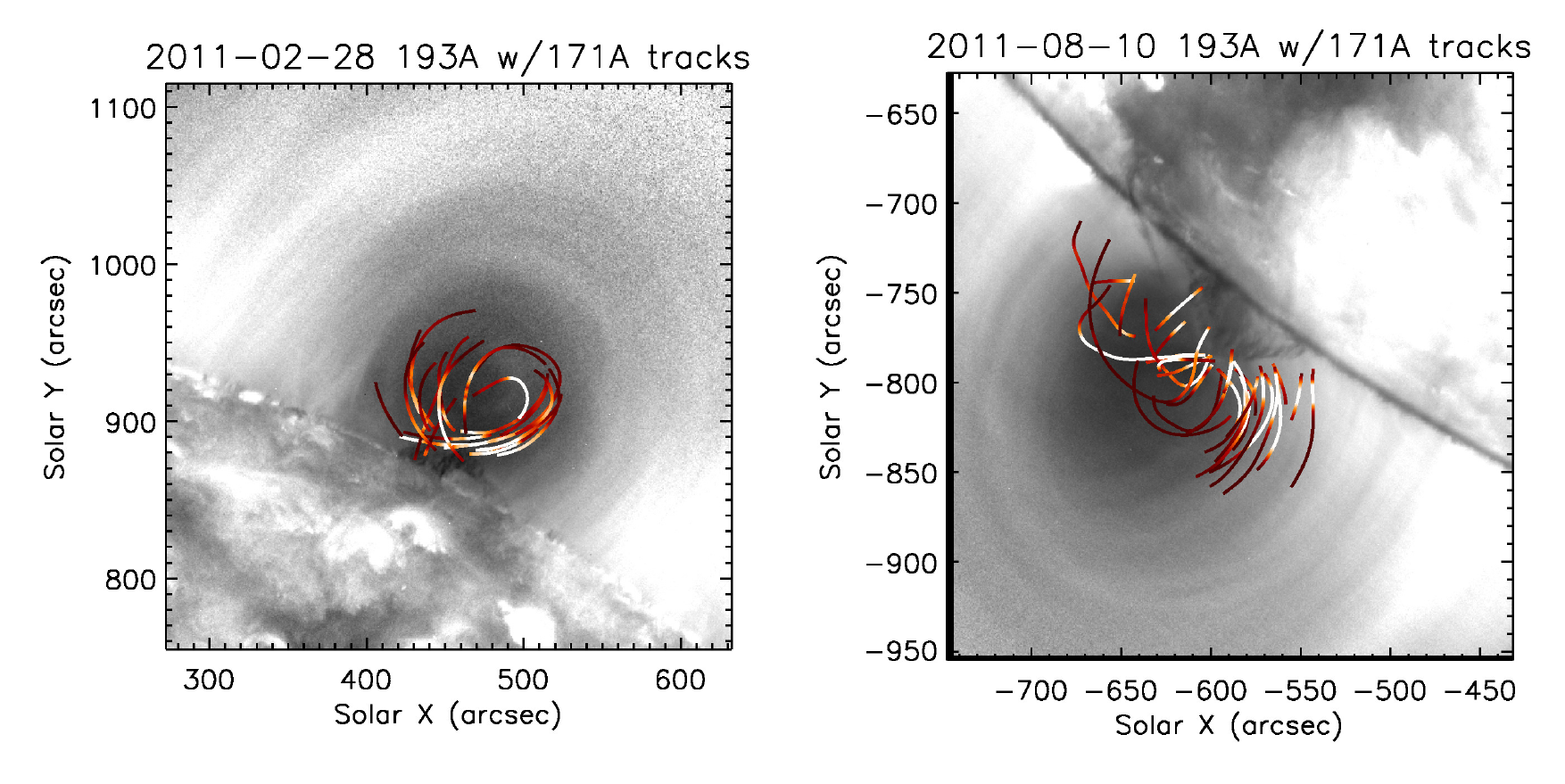}
\caption[Maps of 171\AA~horn structure]{193\AA~map of D1 (left) and D2 (right) in grayscale. Location of horns in the 171\AA~bandpass. Color table advances black to orange to white with increased maximum (in time) contrast to background.Range: 4\% to 50\% \label{fig:o171}}
\end{figure*}
\begin{figure*}
\includegraphics[width=\textwidth]{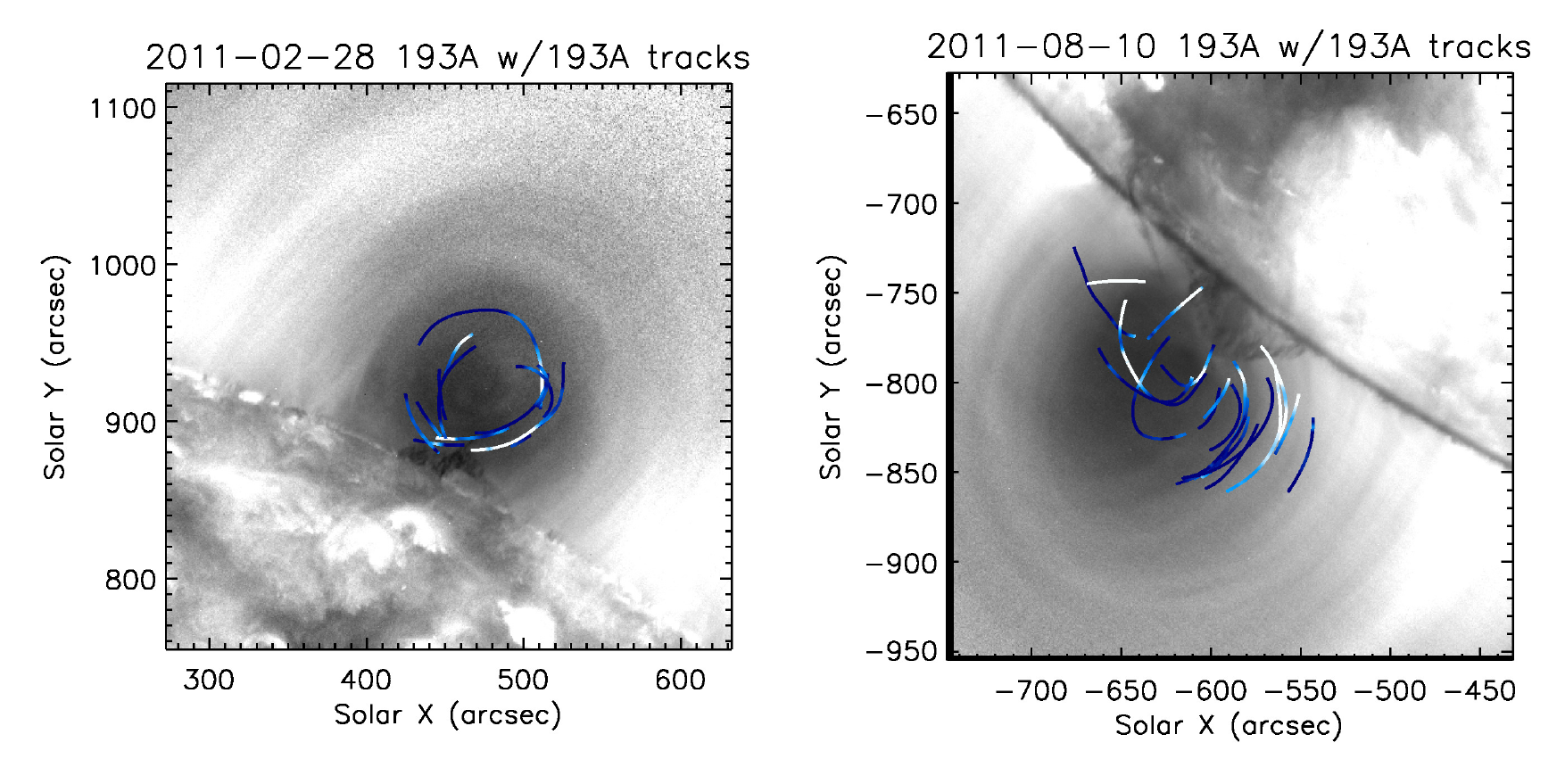}
\caption[Maps of 193\AA~horn structure]{193\AA~map of D1 (left) and D2 (right) in grayscale. Location of brightenings in the 193\AA~bandpass associated with horns. Color table advances black to blue to white with increased maximum (in time) contrast to background. Range: 1\% to 20\%. \label{fig:o193}}
\end{figure*}
\begin{figure*}
\includegraphics[width=\textwidth]{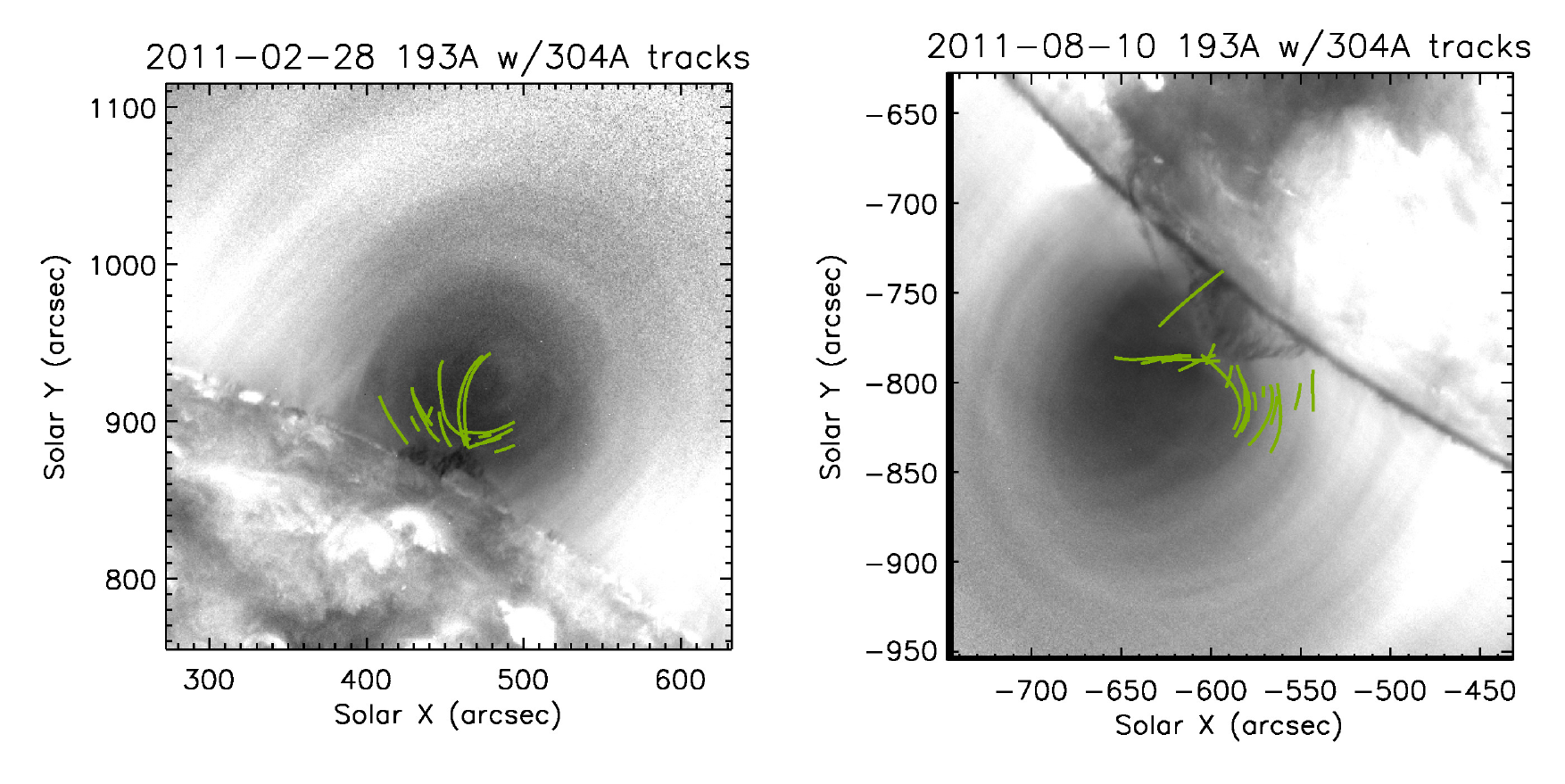}
\caption[Maps of 304\AA~horn structure]{193\AA~map of D1 (left) and D2 (right) in grayscale. Location of prominence extensions in 304\AA~emission associated with the horns. These tracks represent the range of prominence extension or ejection. No color range is applied.\label{fig:o304}}
\end{figure*}
\\\indent
Maps showing the positions and strengths of the 171\AA~horns are shown in Figure~\ref{fig:o171}.
Both maps illustrate the proximity of these coronal features to the prominence.
The dense, absorbing regions of the prominence appear to be the emanation point.
The structure is more complex in D2.
In the 304\AA~movie, we are able to see that there are two unique prominence spines in the D2 dataset.
The smaller northern spine is observed exchanging plasma with the larger southern spine, which is centered under the cavity.
These ``connection'' events have similar traits to the more vertical horns in the 171\AA~light curves despite there different geometries.
However, they do not typically show much structure in 193\AA.
For D1, the overall appearance of the 171\AA~brightenings are concave loop segments surrounding a central axis located at $(+475'',+915'')$ at a height of 1.07 R$_\odot$.\\\indent
Of the 45 horns found in the 171\AA~bandpass, 32 exhibit correlated brightenings in 193\AA~bandpass data.
The associated 193\AA~features are plotted in the maps of Figure~\ref{fig:o193}.
Of the events which do show correlation between 171\AA~and 193\AA~light curves, there is generally not a correlation throughout the entire horn.
The 193\AA~brightenings tend to occur in the upper sections of the 171\AA~horns, the median offset is 6.3 Mm above the 171\AA~base.
The most striking difference between the 171\AA~and 193\AA~features are that the 193\AA~events are significantly dimmer.
The median peak brightness is 7\% in 193\AA~compared to the 34\% in 171\AA.\\\indent
From the 32 joint 193\AA-171\AA~events, we have 450 co-spatial light curves which exhibit correlation.
From these points, we measure an average time lag of -0.6 minutes $\pm 23$ minutes between the peak brightness in 171\AA~and the peak brightness in 193\AA.
The same applies for the event onset time.
This lack of correlation implies that there are both temperature and density perturbations causing the brightenings.
\\\indent
Maps which show the extent of 304\AA~features correlated with 171\AA~horns are shown in Figure~\ref{fig:o304}.
We find 34 of these events.
Unlike Figure~\ref{fig:o171} and~\ref{fig:o193}, there is no color range applied to the 304\AA~features.
We did not extend the same intensity analysis to the 304\AA~dataset because of the large intensity gradients on small-spatial scales within the prominence.
The basis of our comparison with 304\AA~data is changes in the perimeter of the prominence, as was seen in Figure~\ref{fig:close}.
There are generally two type of events we observe in the 304\AA~data.
In 20 of the 304\AA~events, we observe simple extensions of the prominence.
In the other 14 events, we actually observe ejection of prominence material such that no 304\AA~emission is detected between the 304\AA~ejecta and the prominence itself.
In both cases, the median rise time and fall time is 24 minutes.
The median maximum length for ejecta is 10.5 Mm, while the median maximum length for extensions is 7.6 Mm.
In general the rise of 304\AA~material begins after the initiation of the 171\AA~brightening by 30 minutes.
We find a Pearson coefficient of 0.44 for the correlation of the peak 304\AA~extension time and the time of peak 171\AA~brightness.
\\\indent
Velocities are a fundamental quantity associated with dynamics, however this is a difficult measurement to make in prominence horns.
We do not have spectral Doppler observations for these cavities, and the application of feature tracking techniques is not obvious;
our features appear to maintain a constant spatial orientation and position over their 3 hour duration.
As described in Section 3, we have attempted to assess a  velocity based on the time of initial brightening and peak brightening.
While this quantity has physical units, we must be clear on what is being measured in terms of basic physics:
{\it there is a perturbation happening along a coronal loop which is changing the amount of photons being emitted along that loop as a function of time.}
We are measuring the time that a discernible change in photoemission has occurred at a particular position along the loop relative to other positions on the loop.
\begin{figure*}
\includegraphics[width=\textwidth]{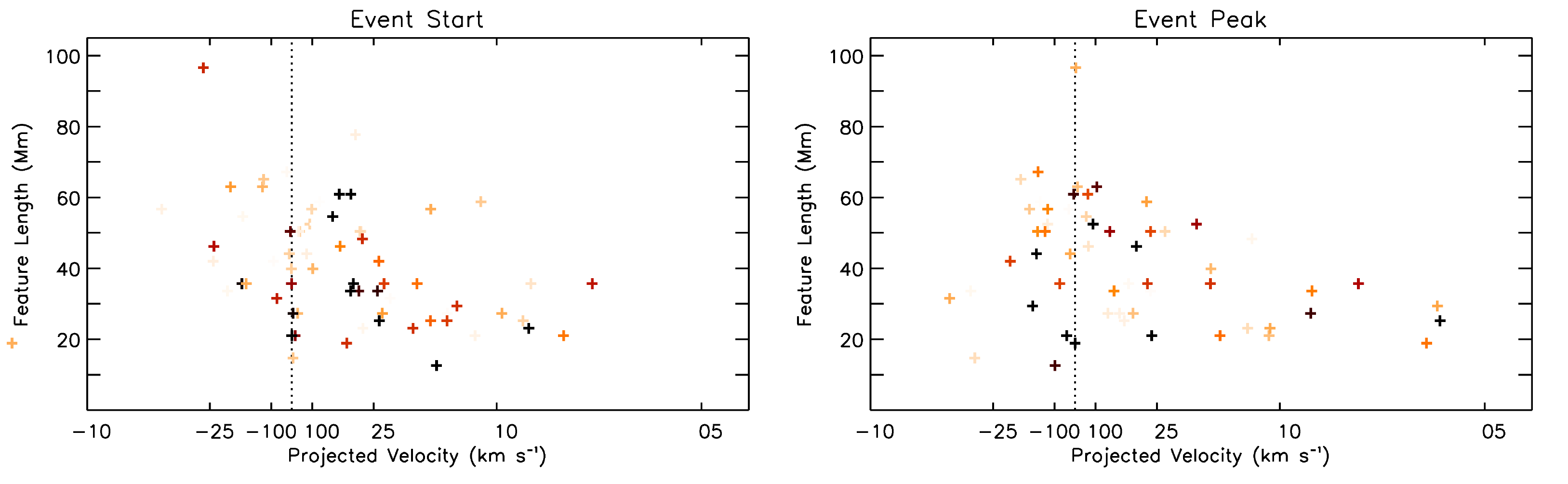}
\caption[Derived velocity of horns]{Scatter plots of the derived horn velocity along the slit versus horn length. Color table advances black to orange to white with increasing $\chi^2$ value. Derived velocity for the horn initiation (left) and horn maximum (right). Black crosses represent the most reliable velocities. $\chi^2$ range for initiation: 0.1 to 1.4;  $\chi^2$ range for maximum: 0.8 to 6.5. The black dotted line represents an infinite and directionless speed ($m=0$).\label{fig:speed}}
\end{figure*}
\\\indent The derived velocities based on the initial brightening and the peak brightening are shown in Figure~\ref{fig:speed} as a function of feature length and reduced $\chi^2$ of the model fit (using Equations 3 and 4).
Feature length is important in this diagnostic because we have more points to track the apparent propagation of the feature.
The peculiar $x$-axis scale is due to the nature of the fitting.
The fit-parameter, $m$ is converted into a velocity by
\begin{displaymath}
v=\frac{\Delta y'}{\Delta t}m^{-1}
\end{displaymath}
where $\Delta y'$ represents the plane of sky distance of 5 slit-pixels (2100 km) and $\Delta t$ is the cadence of the data (360 s).
A negative value of $m$ would imply a propagation from higher radial heights downward, while a positive $m$ implies a propagation from the prominence upward.
The median derived $m$ for the 171\AA~data for the brightness initiation is 0.056 which corresponds to a velocity of 104 km s$^{-1}$.
Given the errors in measurement, we are not able to discriminate between high speed ($|v|>100$ km s$^{-1}$) features moving upward and downward.\\\indent
The derived velocities in Figure~\ref{fig:speed} are sparsely distributed, but the highest quality fits tend to cluster between $|v|>20$ km s$^{-1}$.
These velocities are intriguing considering the event durations.
A transient disturbance traveling at 20 km s$^{-1}$ would advance 60 Mm in 5 minutes.
Our observed features last longer than 1 hour.
This implies that the propagation of the perturbation and the subsequent equilibration or damping of the perturbation are based on different energetic processes and timescales.
\subsection{Estimate of Horn Density}
We wish to estimate the density of the horns, and this requires us to account for projection effects.
Past efforts by \cite{schmit_11} and \cite{kucera_12} have gone through modeling efforts to quantify the density and temperature effects which are manifest in optically-thin EUV emission.
Extending this work to the D1 cavity, we can make a back-of-the-envelope calculation for prominence horns.
Let us consider a line of sight which observes a 53$^\circ$ long cavity embedded with in a 70$^\circ$ long streamer at a latitude of 60$^\circ$.
Using the forward modeling technique of \cite{gibson_10}, we predict there should be approximately $9\times10^{26}$ cm$^{-5}$ of emission measure along the cavity-center line of sight at a height of 1.07R$_s$.
Assuming an isothermal cavity and streamer of 1.7 MK, we can write the intensity ratio of the prominence horn to the static cavity in the form,
\begin{equation}
\frac{I'_{171}}{I_{171}}=\frac{\mathcal{A}_{171}(T=1.7 \mathrm{MK})*\mathrm{EM}_s+\mathcal{A}_{171}(T=T_{peak})*\mathrm{EM}'}{\mathcal{A}_{171}(T=1.7 \mathrm{MK})*\mathrm{EM}}
\end{equation}
where $\mathcal{A}$ represents the AIA temperature response function per unit emission measure.
The forward model predicts that most of the ambient 171\AA~emission comes from low-density, weakly-emitting plasma, but the plasma is quite extended (the full-width half-max for emissivity distribution is $\pm 70$ Mm out of the plane of the sky).
The ratio of $\mathcal{A}_{171}(T=1.7~\mathrm{MK})/\mathcal{A}_{171}(T=T_{peak})$ is approximately 4.5\%.
We have divided emission measure distribution into three components
\begin{displaymath}
\mathrm{EM}=\int_{0}^{\infty}n^2~d\ell,
\end{displaymath}

\begin{displaymath}
\mathrm{EM}_s=\int_{3~\mathrm{Mm}}^{\infty}n^2~d\ell,
\end{displaymath}

\begin{displaymath}
\mathrm{EM}'=\int^{3~\mathrm{Mm}}_{0}n'^2~d\ell,
\end{displaymath}

where $n'$ is the horn density and $n$ is radial density profiles from \cite{schmit_11}.
These emission measures are the full line of sight, the horn region in the ambient cavity, and the horn region during the 171\AA~brightening respectively. 
We estimate the line-of-sight width of the horn at 6 Mm.
Substituting in these equations, we can solve the minimum density perturbation necessary to produce a 50\% increase in 171\AA~intensity
\begin{displaymath}
\frac{n'(\ell=0)}{n(\ell=0)}\approx 1.2
\end{displaymath}
This increase in density assumes that the plasma has also cooled to 1 MK, where it emits 18 times more strongly than at 1.7 MK, the original imposed temperature.
This assumption makes it such that the predicted density change is a minimum value, whereas a weaker temperature perturbation would result in a weaker enhancement in emission.
Overall, we predict that a 6 Mm wide swath of plasma would have to increase its emission by a factor of 22 to produce an overall brightening of 50\% when accounting for background emission.
\\\indent
We can apply this same technique for 193\AA~emission where we use the same temperatures and the derived $n'$ and solve for $I'_{193}/I_{193}$.
We now have the ratio in the temperature response function of
\begin{displaymath}
 \frac{\mathcal{A}_{193}(T=1.7~\mathrm{MK})}{\mathcal{A}_{193}(T=1~\mathrm{MK})}=2.4
\end{displaymath}
The reduction in temperature which raised the emissivity for the 171\AA~bandpass by a factor of 20, acts to reduce the 193\AA~emissivity.
The predicted ratio $I'_{193}/I_{193}$ has a value 0.98 compared to 1.5 for $I'_{171}/I_{171}$.
In other words, the intensity produced by a  corona dominated by $>1.5$ MK structure will not vary drastically for localized (short line of sight contributions) cool-coronal events.
\section{Conclusions}
We have demonstrated that the cavity exhibits fundamentally stronger dynamics than the surrounding streamer.
This was illustrated through a statistical analysis of EUV observations from the SDO/AIA instrument.
In particular, the statistics indicate that the lower-cavity region surrounding the prominence undergoes strong brightening events in the 171\AA~bandpass, which is typically dominated by 1 MK plasma.\\\indent
Analyzing movies in the 171\AA~bandpass, the observations indicate that these events are coronal loop segments, called prominence horns, which extend from the prominence into the corona.
Horns project into the interior of the cavity.
The 171\AA~emission of horns is often correlated with weakly-enhanced emission in 193\AA~bandpass, which contains hotter coronal emission.
The 171\AA~emission is often co-spatial with extensions of the prominence material as seen in the 304\AA~bandpass.\\\indent
We have gone through many details of these emission signatures, but we have not yet addressed the physical nature of these events.
There are several obvious puzzles present in the data.
First, how do we explain the apparent co-spatial 304\AA~and 171\AA~emission?
There is a large temperature discrepancy between the theoretical peak emissivity of He II 304.3\AA~and Fe IX 171.4\AA.
There are two possible explanations: cool blends in the 171\AA~bandpass or projection effects.
\\\indent
There are cool blends in the red periphery of the 171\AA~bandpass, in particular resonance lines of O V 172.2\AA~and O IV173.0\AA.
\cite{delzanna_11} concluded thorough analysis of Hinode/EIS data that the contributions of these lines is limited to a few percent on disk.
These lines will not emit strongly in equilibrium because of the large difference between the excitation energy (70 eV) and the temperature associated with the peak ion population (O VI population peaks at 4$\times10^5$ K $\approx$ 35 eV).
Radiative recombination could play a role in populating these excited levels in a non-equilibrium processes.
However, the magnitude and duration of this recombination emission is unlikely to explain prominence horns.
\\\indent
Another possible explanation for the correlated 304\AA~and 171\AA~emission is projection effects.
Horns may have a significant component of their geometry aligned along the line of sight.
Emission which appears cospatial would occur along separate segments of an individual magnetic field line.
The prominence-corona transition region, which would separate these segments, is predicted to be less than 5 Mm thick.
\\\indent
Another peculiar aspect of prominence horns is the weak correlation between 171\AA~emission and 193\AA~emission.
There are relative weak separations between the ionization energies for Fe IX and Fe XII.
This has the effect that the predicted AIA temperature response function has significant overlap between the 171\AA~and 193\AA~bandpasses.
Moreover, there is precedence that Fe IX 171\AA~and Fe XII 195\AA~emission are often highly correlated in dynamics heating events \citep{viall_11,aschwanden_00}.
The most likely explanation for the weak correlation in horns is that the dominant component emission in the 171\AA~is formed at temperatures less than 0.8 MK.
This cool plasma would not have strong emission in the 193\AA~bandpass.
\\\indent
In Section 2, we discussed that there was more variability in the 171\AA~bandpass than the 193\AA~bandpass.
While horns explain our statistical analysis of the 171\AA~cavity, the associated 193\AA~component of the horns does not explain the variability in that bandpass.
In examining the light curves, we find that there are intensity variations a factor of 3 larger in the 193\AA~cavity than the intensity variations related to the horns.
The power seems to be strongest for variations with a 4-8 hour period and is more evenly distributed between dimming and brightening events.
We believe that it is this variability component that \cite{wang_stenborg} have studied through their wavelet analysis.
\\\indent
We have created a framework with which we could measure the characteristic velocity of the prominence horns, but found that the spread in the derived values was large relative to the values themselves.
We find several examples from our 45 events where an entire 50 Mm long structure undergoes a uniform and simultaneous change in intensity.
In terms of basic physics, these large-scale isotropic changes are most readily explained through thermal conduction.
\\\indent
Conduction is a diffusive operator.
Given a thermal perturbation on a coronal loop, anisotropic thermal conduction rapidly redistributes the energy associated with the perturbation.
These characteristics match the scale ($>$ 50 Mm) and velocity ($>50$ km s$^{-1}$) of horns. 
In the corona  however, thermal changes and density changes are inseparable.
Due to the different velocity scales associated with thermal conduction and siphon flows, a change in temperature will always result in a change in density.
A change in density, given mass conservation, requires velocities.
The hydrodynamic equations must be self-consistently solved to truly understand the implications for any thermodynamic perturbation.
Of special interest to these observations is a detailed analysis of a catastrophic cooling model for prominence formation.
A followup paper is underway comparing the light curves of prominence horns with the thermal non-equilibrium model of \cite{karpen_06}.
\\\indent
We have not yet discussed the observed geometry of the prominence horns.
Horns are tightly-collimated and exhibit little structural variation over their duration.
These characteristics indicate that these are magnetically-aligned structures, or more specifically segments of coronal loops.
We have speculated they may be explained as thermal structures, and thermal conduction is dominantly field aligned.
Thus, we posit the extracted prominence horns are indicative of the magnetic field geometry of the lower cavity.
\\\indent
In both D1 and D2, a discernible axis can be found which the horns circumscribe.
Horns imply that there is a component of the prominence magnetic field which is non-axial.
This is an indication that twist is present in the quiescent prominence-cavity system.
However, we cannot be completely certain that these magnetic structures are within the cavity itself.
It remains possible that horns and the cavity inhabit separate field lines which overlay each other in optically thin projection.
It is also interesting that despite the relative symmetry of horns about the prominence spine, we do not find evidence that horns on one side are correlated with horns on the other (we do not observe dynamics which extend through the prominence spine).
\\\indent
These observations are the first attempt to statistically quantify the connection between cavity dynamics and prominence dynamics.
We have identified a strong emission variability connection between the prominence and the cavity, but we have not gone so far as to determine the physical source of these dynamics.
There is a compelling paradigm for the prominence-cavity connection:  a mass exchange feeds condensing plasma out of the low-density cavity into the cool, overdense prominence.
The models for catastrophic cooling make predictions on what coronal perturbations are necessary to drive this process.
In the forthcoming article, we will test these predictions against the emission variability we have documented in the prominence environment.

%\bibliographystyle{apj.bst}

%\bibliography{mybibliography}

\end{document}